\documentstyle[12pt,aaspp4]{article}

\lefthead{Sankrit, Sembach, Canizares} \righthead{M87 Ly$\alpha$ Emission}

\begin{document}


\title{HST-FOS Observations of M87: Ly$\alpha$ Emission
from the Active Galactic Nucleus\altaffilmark{1}}

\author{Ravi Sankrit\altaffilmark{2}, 
Kenneth R. Sembach\altaffilmark{2},  
Claude R. Canizares\altaffilmark{3}}

\altaffiltext{1}{Based on observations with the NASA/ESA {\em Hubble 
Space Telescope,} obtained at the Space Telescope Science Institute, 
which is operated by AURA, Inc., under NASA contract 5-26555.}

\altaffiltext{2}{Department of Physics and Astronomy, 3400 N. Charles St.
The Johns Hopkins University, Baltimore, MD 21218}

\altaffiltext{3}{Center for Space Research, MIT, Cambridge, MA 02139}


\begin{abstract}
The Faint Object Spectrograph on the Hubble Space Telescope was used to
obtain spectra of the central region of M87.  These spectra cover the
wavelength range 1140~\AA\ -- 1606~\AA\ and have a resolution of $\sim$
1~\AA.  The nuclear continuum is clearly visible in the spectra.  The
only strong line that is observed is Ly$\alpha$, which has a velocity
width of about 3000 km~s$^{-1}$ (higher than the width of any line
previously observed in the M87 nucleus).  There is also a marginal
detection of C~IV $\lambda$1549.  The ratio of Ly$\alpha$ to C~IV in
the nuclear spectrum is at least a factor of 2 higher than in a
spectrum taken at a position on the disk $\sim$~0\farcs6 away from the
nucleus by Dopita et al.  This enhancement of Ly$\alpha$ at the nucleus
could point to significant differences in the properties of the
emitting gas and/or the excitation mechanism between the outer and
inner disk regions.  The strength of the observed Ly$\alpha$ places
limits on the properties of the absorbing gas present within M87.  For
instance, if the hydrogen column at the systemic velocity of M87 is
greater than about 10$^{18}$ cm$^{-2}$ then it can cover only a small
fraction of the line emitting region.  Spectra separated by 5 days show
a 60\% difference in the Ly$\alpha$ flux, but the same continuum
level.  This could be due to either a displacement between the aperture
positions for the two sets of observations, or it could be due to
intrinsic variability of the source.  The current observations do not
strongly favor either of these alternatives.  The observations do
show, however, that the Ly$\alpha$ line is a useful tracer of
kinematics in the M87 nucleus.

\end{abstract}

\keywords{galaxies: active --- galaxies: individual(M87) --- 
galaxies: nuclei}


\section{Introduction}

The giant elliptical galaxy M87 (NGC~4486), at the center of the Virgo
cluster with its spectacular jet, is a prototypical active galactic
nucleus.  Its nucleus contains a luminous accretion disk, discovered in
a {\it Hubble Space Telescope} (HST) WFPC2 H$\alpha$ image
(\cite{for94}).  Spectroscopy of the region using the HST {\it Faint
Object Spectrograph} (FOS) showed that the gas in the disk is in
Keplerian rotation about a mass concentration of $\sim2.5\times10^9$
M$_{\sun}$ within the inner 0\farcs25 of the nucleus (\cite{harms94}).
These two companion papers provided strong evidence for the existence
of a supermassive black hole at the center of M87, something first
suggested several years ago (\cite{you78}, \cite{sar78}).  Subsequent
investigations of the disk kinematics have strengthened the case for
such a black hole (\cite{mar97}, \cite{mac97}).

More recently, based on the variability of the nuclear continuum, the
continuum spectral shape, strong relativistic boosting, and the
detection of superluminal motions in the jet, it has been suggested
that M87 could be a misaligned BL Lac object (\cite{tsv98a}).  FOS
observations of the disk away from the nucleus show a LINER spectrum
with evidence for shock heated gas (\cite{dop97}).  The luminosity of
the M87 nucleus at all wavelengths is significantly lower than the
luminosity of typical radio loud quasars such as 3C\,273, which led
\cite{rey96} to suggest that the M87 black hole accretes in an
advection-dominated mode.  Properties of the M87 nucleus and disk have
been reviewed recently by \cite{for98}, \cite{tsv98b} and
\cite{tsv98c}.

In this paper we present new FOS observations of the M87 nucleus.
These spectra differ from earlier spectra of M87 taken with the same
instrument -- they are centered on the nucleus, they have higher
spectral resolution, and they cover the ultraviolet wavelength range.
In the following section we describe the observations.


\section{Observations}

The observations presented in this paper were taken as part of HST
program GO-5921 (P.I. Canizares).  Spectra of the M87 nucleus (R.~A.
12$^h$30$^m$49$^s$.43, Dec. +12$^\circ$23\arcmin27\farcs9 J2000) were
taken through the 0\farcs26 diameter circular aperture, using the G130H
grating, which covers the wavelength region 1140~\AA\ -- 1606~\AA.  The
spectra have a resolution of about 1~\AA, corresponding to $\sim$ 250
km~s$^{-1}$ at the rest wavelength of the Ly$\alpha$ line
(1215.7~\AA).  The data were reduced with the standard HST pipeline
software available at the time of the observations.  Inspection of the
zero-levels of the spectra near the Ly$\alpha$ line showed that there
are no systematic background errors in the reduced data.

An initial set of observations obtained in May 1996 had a pointing
error -- the aperture was centered more than 1\arcsec\ away from the
nucleus.  A second set of observations was obtained in January 1997
with the correct pointing.  Five spectra having a total exposure time
of 9970 seconds were taken on January 18, and six spectra  having a
total exposure time of 12,720 seconds were taken on January 23.  In the
following discussion, unless explicitly mentioned otherwise, we will be
referring to the summed spectrum from the second set of observations.
When needed, we will refer to the May 1996 observations as ``epoch 1'',
and distinguish between the January 18 and January 23, 1997
observations by referring to them as ``epoch 2a'' and ``epoch 2b'',
respectively.  The observations are preserved in the HST archive with
identifications Y3170104T--9T (epoch 1), Y3175106T--AT (epoch 2a), and
Y3175202T--7T (epoch 2b).

In the following sections we present our results.  Continuum emission
from the nucleus is detected at levels consistent with previous
observations.  Ly$\alpha$ is the only emission line that is clearly
detected, though there is a marginal detection of \ion{C}{4}
$\lambda$1549.  The Ly$\alpha$ flux in the epoch 2b spectrum is
significantly higher than in the epoch 2a spectrum.  The difference is
due either to an offset in the aperture positions at the two epochs,
which leads to sampling of slightly different regions of the nucleus,
or to intrinsic temporal variability in the source on a 5 day time
scale.


\section{The Nuclear Continuum}

In Figure \ref{CONT} we plot the observed flux ($\nu$F$_{\nu}$) against
frequency between $2.0\times10^{15}$ Hz and $2.4\times10^{15}$ Hz
(1500~\AA\ -- 1250~\AA) for both epoch 1 (dotted line) and epoch 2
(solid line).  The epoch 1 observations were centered 1\farcs4 away
from the nucleus, which at the distance of M87 (15 Mpc, following H94)
corresponds to about 100 pc.  The continuum at this location is
essentially zero.  The average epoch 2 value $\nu$F$_{\nu} \approx
1.2\times10^{-12}$~erg~s$^{-1}$~cm$^{-2}$ is the same as that presented
by \cite{tsv98b} and is also consistent with the value obtained by H94
at $7\times10^{14}$~Hz (see Table 1 of \cite{rey96}).  The wavelength
range of our spectrum is too limited to derive an accurate spectral
index for the emission.


\section{Line Emission}

In Figure \ref{LYA} we plot the spectrum of the M87 nucleus between
1200~\AA\ and 1250~\AA\ (solid line).  The Ly$\alpha$ emission line is
clearly separated from the geocoronal emission (expectedly, since the
systemic velocity of the galaxy is about 1300 km~s$^{-1}$; H94).  The
dotted and dashed lines show two estimates of the absorption due to
Galactic hydrogen as a function of wavelength.  (The transmitted
fraction can be read off the numbers on the y-axis scale directly).
The dotted curve was derived from the Galactic \ion{H}{1} 21cm emission
profile presented by \cite{dav75}, which when integrated yields a total
column N(\ion{H}{1}) $\approx$ 3.4$\times$10$^{20}$ cm$^{-2}$.  Their
21cm profile shows a broad emission wing from $\approx$ +20 to +60
km~s$^{-1}$. We checked their result using the stray-radiation
corrected Leiden/Dwingeloo survey (\cite{har97}).  We derived an
emission profile in the direction of the M87 nucleus by averaging 8
spectra obtained at points surrounding the nucleus at an angular
distance of 1\arcdeg.   The positive velocity wing is not present in
either the individual or averaged stray-corrected spectra.  We find an
integrated N(\ion{H}{1}) $\approx$ (2.1$\pm$0.3)$\times$10$^{20}$
cm$^{-2}$, or about 62\% of the Davies \& Cummings (1975) value. The
maximum N(\ion{H}{1}) in any of the 8 spectra is 2.6$\times$10$^{20}$
cm$^{-2}$.  We show the corresponding absorption profile as a dashed
curve in Figure \ref{LYA}.  The exact shape of the Ly$\alpha$
absorption is less sensitive to the details of the velocity
distribution than to the total value of N(\ion{H}{1}).  One effect of
the Galactic absorption is that the observations provide only a lower
limit to the intrinsic Ly$\alpha$ emissivity of the source.  The
apparent line centroid of the M87 emission may be skewed to slightly
higher velocities because the Galactic absorption lies on the negative
velocity side of the M87 Ly$\alpha$ profile.

The observed Ly$\alpha$ emission profile is very wide, extending from
1218~\AA\ to 1230~\AA.  This implies a total velocity width of about
3000 km~s$^{-1}$.  The peak flux occurs at 1222~\AA, which corresponds
to a velocity of about 1480 km~s$^{-1}$.  This is consistent with the
systemic velocity of M87 (1309 km~s$^{-1}$; H94) since the spectral
resolution is only 250 km~s$^{-1}$.  The observed flux, integrated
between 1218~\AA\ and 1230~\AA\, is $\sim (8.9 \pm
0.1)\times10^{-14}$~erg~s$^{-1}$~cm$^{-2}$.  If we assume that the
emission fills the 0\farcs26 diameter aperture, this implies a surface
brightness of $\sim
1.7\times10^{-12}$~erg~s$^{-1}$~cm$^{-2}$~arcsec$^{-2}$.

The Ly$\alpha$ emission profile (Figure \ref{LYA}) also exhibits a
shape discontinuity at $\sim$1226~\AA.  In order to quantify the
velocity width of the profile in a more formal manner, we created an
artificial line profile by reflecting the observed line profile to the
longer wavelength side about the observed peak.  We then fit the
resulting profile with the sum of two gaussians with FWHM equal to 1525
km~s$^{-1}$ and 3490 km~s$^{-1}$, each contributing roughly half the
flux.  The narrower component has a velocity width comparable to those
found by H94 for the optical lines in their FOS spectrum of the
nucleus.  The broad component has a line width higher than has been
observed in optical emission lines.  We emphasize that the velocity
widths depend in part on the assumption that the broad and narrow
components are centered at the same wavelength and have symmetric
velocity distributions about this wavelength.

We estimate the intrinsic Ly$\alpha$ intensity of the M87 nucleus by
correcting the observed flux for the Galactic absorption shown in
Figure \ref{LYA}.  For absorbing column densities of
$2.1\times10^{20}$~cm$^{-2}$ and $3.4\times10^{20}$~cm$^{-2}$, the
integrated fluxes between 1218~\AA\ and 1230~\AA\ are
$\sim~12.8\times10^{-14}$~erg~s$^{-1}$~cm$^{-2}$ and
$\sim~16.2\times10^{-14}$~erg~s$^{-1}$~cm$^{-2}$ respectively.  Between
30\% and 45\% of the Ly$\alpha$ is absorbed by Galactic hydrogen.  
The location of the peak in the corrected profile is very sensitive
to the absorbing column used because of the large corrections
shortward of 1219 \AA.  For a two component fit, allowing the fit
parameters to vary for both components simultaneously, the centroids
are ill-constrained.  If we repeat the exercise of fitting a symmetrized
profile, then for the case of the lower absorbing column, the broad
component contributes about one-fourth the intensity.  (The broad component
contributed half the flux in the fit to the observed profile).

The only other line we detect in our spectrum is the \ion{C}{4}
1549\,\AA\ line.  In Figure \ref{C4} we plot our spectrum between
1500~\AA\ and 1600~\AA.  The solid line is the data rebinned over 4
pixels (about 1~\AA).  The dotted line shows the unsmoothed data to
give an idea of the noise level in the original spectrum.  Overplotted
on the data are two gaussians (dashed lines) added to a linear
background, to simulate the ``expected'' \ion{C}{4} flux, which will be
discussed next.  Both gaussians have a FWHM of 15.0~\AA\ ($\sim$ 3000
km~s$^{-1}$) and are centered at 1555.8~\AA\ ($\sim$1300 km~s$^{-1}$).

Both Ly$\alpha$ and \ion{C}{4} $\lambda$1549 were observed by D97 in
their spectrum of the disk at the center of M87.  (Their aperture was
placed so as to avoid the nucleus).  In their spectrum the ratio of
observed fluxes $F_{1549}/F_{Ly\alpha} \approx 0.2$.  Applying this
ratio to our Ly$\alpha$ measurement, we estimate a \ion{C}{4}
$\lambda$1549 flux of $1.8\times10^{-14}$~erg~s$^{-1}$~cm$^{-2}$.  The
stronger gaussian in Figure \ref{C4} is constructed to have this flux,
and in spite of the large velocity width, it is clearly incompatible
with the observations.  Thus, the observed \ion{C}{4} to Ly$\alpha$
ratio is weaker in the nucleus than in the disk.

We can obtain the ``expected'' \ion{C}{4} flux in another way, using
the H$\beta$ flux instead of Ly$\alpha$.  H94 observed the H$\beta$
flux from the nucleus to be about
$2.5\times10^{-15}$~erg~s$^{-1}$~cm$^{-2}$ and in the D97 spectrum, the
ratio $F_{1549}/F_{H\beta} \approx 3.6$.  Using this ratio, we estimate
a \ion{C}{4} $\lambda$1549 flux of
$9\times10^{-15}$~erg~s$^{-1}$~cm$^{-2}$, which is half the value
derived using Ly$\alpha$.  The weaker gaussian in Figure \ref{C4} is
constructed to have this flux, and it is a reasonable upper limit on
the observed \ion{C}{4} emission.  (Note that since the data are noisy,
we do not attempt to find a ``best fit'' spectrum to the \ion{C}{4}
line profile).

In our spectrum of the M87 nucleus, we observe $F_{1549}/F_{Ly\alpha} \le 0.1$
(see the discussion above and Figures \ref{LYA} and \ref{C4}).  This is
at least a factor of 2 lower than the value obtained by D97 for their
disk spectrum.  Furthermore, the Ly$\alpha$ emission in our spectrum
(Figure \ref{LYA}) extends out to higher velocities (relative to line
center) than previously reported for any line.  As discussed above, if
we assume that the observed Ly$\alpha$ line is symmetric, it can be fit with two
components each containing about half the flux.  This makes the observed ratio
of \ion{C}{4} to ``narrow'' Ly$\alpha$ compatible with the value obtained by D97.
Therefore, one interpretation of our spectrum is that the narrower
Ly$\alpha$ emission and all of the \ion{C}{4} emission comes from part
of the disk which has a LINER spectrum like that observed by D97, and
the remaining Ly$\alpha$ (the broad component) arises in a region with
different characteristics (which may or may not be part of the disk).
In this picture, the first component is due to a fast shock such as the
one invoked by D97 to explain their spectrum.  The second component
could be due to slower shocks ($v_{shock} \lesssim$ 100 km s$^{-1}$)
which do not produce \ion{C}{4}, or it could be due to photoionization
of material by the nuclear continuum.  

It is worthwhile pointing out that the low value of \ion{C}{4} to
Ly$\alpha$ does not rule out photoionization as a possible excitation
mechanism for the putative second component.  The M87 nucleus is
underluminous compared to typical AGN (\cite{rey96}) and the incident
flux on any photoionized gas is correspondingly low.  To address this
issue, we used the M87 spectrum presented by \cite{rey96} to
calculate simple slab geometry photoionization models using CLOUDY
(\cite{fer98}).  We find that the ionization parameter is $10^{-3}$ for
a cloud with density n$_{H} = 10^{6}$ cm$^{-3}$, lying about 1 pc away
from the nucleus.  For this value of the ionization parameter, the
model predicts $F_{1549}/F_{Ly\alpha} \sim 0.01$.  Clearly such
weak \ion{C}{4} emission would not be seen in our spectrum.  We note
that 1 pc corresponds to about 0\farcs015 at the distance of M87, and
is about 1/100 the radius of the ionized disk (\cite{for98}).

In order to distinguish between shock excited gas and photoionized gas
more diagnostic lines need to be measured.  Based on our present data,
we conclude that the fast shock which produces the disk spectrum of D97
does not explain all the Ly$\alpha$ emission.


\section{Absorbing Gas within M87}

Absorption line systems for both our Galaxy and M87 have been observed
by \cite{tsv98b}.  Their near-ultraviolet and optical spectra  contain
broad ($\sim 400$ km~s$^{-1}$) absorption lines arising in M87 from
neutral and low ionization species (\ion{Na}{1}, \ion{Mg}{1},
\ion{Ca}{2}, \ion{Mg}{2}, \ion{Mn}{2}, \ion{Fe}{2}).  They suggest a
simple model of the nucleus in which the line of sight to the central
point source passes through an outflow from the inclined, turbulent
disk.

Higher resolution ($\sim$10 km~s$^{-1}$) optical absorption line
observations of \ion{Na}{1} and \ion{Ca}{2} (\cite{car92};
\cite{car97}) reveal at least two, and perhaps more, cool interstellar
clouds in the nuclear region of M87.  The absorption is spread over at
least two components at +990 and +1297 km~s$^{-1}$, each with an
equivalent \ion{H}{1} column density of $\ge~5\times10^{17}$
cm$^{-2}$ and an internal velocity dispersion of $<$~7.5 km~s$^{-1}$.
They interpret these lines as coming from a population of clouds with
radii of $10^{12}$ cm -- $10^{16}$ cm, which covers only a fraction
(perhaps 10\%) of the nuclear region responsible for the continuum.
High spatial resolution (beam width $\approx$ 1\arcsec) VLA
observations of the \ion{H}{1} 21\,cm absorption in the core of M87 by
\cite{vang89} yield an optical depth of $<$~0.007 and
N(\ion{H}{1})\,$<$\,5.1$\times$10$^{19}$ cm$^{-2}$. There is some
debate about whether the total \ion{H}{1} column density may in fact be
larger than this value, but high-resolution observations of species
other than trace ions such as \ion{Na}{1} and \ion{Ca}{2} will be
needed to make an accurate assessment of the amount of gas present and
its detailed velocity distribution.  For the purposes of the discussion
that follows, we will consider a range of values of N(\ion{H}{1}) for
the M87 gas.

We note that our FOS data are not of sufficient spectral resolution and
S/N to make meaningful determinations of the metal-line strengths in
the M87 gas. For an \ion{H}{1} column density of 5 $\times$ 10$^{19}$
cm$^{-2}$ and a covering fraction of unity, even the strongest metal
lines in this wavelength range (e.g., \ion{C}{2} $\lambda$1334.5,
\ion{O}{1} $\lambda$1302.2) would have apparent absorption depths of
only $\approx$\,10--25\%, relative to the continuum, at this
resolution. This is less than the 1$\sigma$ scatter of the noise about
the mean continuum level.

It is possible, however, to qualitatively assess the effects the
\ion{H}{1} associated with the known interstellar clouds in M87 might
have on the observed Ly$\alpha$ emission from the nucleus.  In Figure
\ref{ABS} we plot our epoch 2 spectrum between 1210~\AA\ and 1235~\AA.
Also plotted are the Ly$\alpha$ absorption profiles (again as
transmittances) which we have  modeled using the velocity structure for
the absorbers seen by \cite{car97}.  We have used hydrogen column
values, N(\ion{H}{1}) = $1~\times~10^{18}$ cm$^{-2}$ and
$5~\times~10^{19}$ cm$^{-2}$.  From Figure \ref{ABS} it is clear that
for the higher N(\ion{H}{1}) case, most of the Ly$\alpha$ emission
would be suppressed due to hydrogen absorption.  Perhaps equally
important, photoelectric heating of dust grains present in cool
absorbing gas would efficiently destroy Ly$\alpha$ photons
(\cite{spi78}).  Yet we detect strong Ly$\alpha$ emission with little
evidence of a sharp discontinuity in the profile at the velocities of
the M87 absorbers.  This implies that the absorbing gas covers only a
fraction of the line emitting region.  More information is needed about
the velocity structure and distribution of gas along the sight line to
adequately model the radiative transfer of resonance line photons
escaping the nuclear regions of M87.

Spectral fits to moderate resolution X-ray data from M87 with the solid
state detector on the Einstein Observatory gave evidence for the
presence of excess absorption, in the central parts of this very
extended source (\cite{whi91}). If attributed to material with cosmic
abundances, the implied excess hydrogen column density was
$15~\times~10^{20}$~cm$^{-2}$.  By analogy to other clusters, this
absorption was associated with the apparent cooling of the X-ray
emitting gas.   An ASCA observation by \cite{mat96} failed to detect
such large absorption but suggested the presence of an excess column
density of $\sim~2.5~\times~10^{20}$~cm$^{-2}$.  Both these values are
clearly ruled out along the line of sight to the nucleus by our
observations (Figure \ref{ABS}), if the hydrogen is neutral.  Since
hydrogen has a negligible contribution to the X-ray absorption cross
section, we cannot rule out a partially ionized absorber.


\section{Ly$\alpha$ Flux Variations}

In the top panel of Figure \ref{LYAVAR} the epoch 2a spectrum (solid
line) and the epoch 2b spectrum (dotted line) are plotted between
1200~\AA\ and 1250~\AA.  It is clear from the plot that the Ly$\alpha$
flux is significantly higher in the later observation.  The observed
fluxes integrated between 1218~\AA\ and 1230~\AA\ (cf. \S4 above) are
$6.6\times10^{-14}$~erg~s$^{-1}$~cm$^{-2}$ and
$10.7\times10^{-14}$~erg~s$^{-1}$~cm$^{-2}$ for epochs 2a and 2b
respectively, corresponding to a 60\% increase.  In the bottom panel,
the ratio of flux in the epoch 2b spectrum to the flux in the epoch 2a
spectrum is plotted for this narrower wavelength range.  The epoch 2b
flux is higher over the entire line, with the ratio jumping from
$\sim$1.3 to $\sim$2.4 at 1223~\AA.  Also shown in the figure are 
plots of the fluxes in arbitrary units.  The plots of fluxes as well as the
ratio have been binned over 4 pixels for clarity.

There are two possible explanations for the flux differences observed.
First, the aperture locations for the two observations could have been
offset in such a way as to cause the observed fluxes to be different.
Second, the source may actually be variable on a time scale of $\sim$5
days.  Both alternatives are interesting, though we cannot confidently
distinguish between them without additional observations.

\subsection{\it{Offset in Aperture Location}}

The flux from the center of M87 is dominated by the nucleus.  However,
the existence of an extended emitting disk makes it somewhat difficult
to center the 0\farcs26 FOS aperture exactly on the nucleus.  The
target was acquired for the epoch 2a observations using a series of
ACQ/PEAK raster scans (\cite{voi97}).  For epoch 2b, the earlier epoch
2a co-ordinates were ``re-acquired'' and a single scan was performed to
find the peak.  We examined the ACQ/PEAK data and estimated that the
offset in aperture positioning could be as much as 0\farcs1 between the
two epochs.  However, from the actual counts in the ACQ/PEAK data, and
more importantly because the continuum flux levels are the same in both
epoch 2a and epoch 2b spectra, we believe that the nucleus itself was
included in the 0\farcs26 diameter aperture in both cases.  This is
supported by the fact that we find the same continuum flux in our
spectra that \cite{tsv98b} found in their low resolution FOS spectrum
of the nucleus.

The center of M87 is a very complicated region where, in addition to
the black hole and Keplerian disk, there are a number of optical
filaments which are thought to originate in a wind from the disk.
Large non-circular velocities have also been observed at several
locations within 0\farcs2 ($\sim$ 14 pc) of the nucleus.  (See
\cite{for98} for a review of these and other properties of the
region).  Therefore, an aperture displacement of 0\farcs1 between two
observations could lead to a significant change in the observed
Ly$\alpha$ flux.  At the same time, the velocity distribution of the
emitting regions must be such that it reproduces the increase in flux as a
function of wavelength over the entire line (Figure \ref{LYAVAR},
bottom panel).  Future studies with accurate aperture (or slit)
positioning will be able to use the Ly$\alpha$ line to map out the
kinematics of the nuclear region.  We note that the estimated
positioning error of 0\farcs1 corresponds to about 7 pc at the distance
of M87 (which we take to be 15 Mpc).  This is $\sim
6\times10^{4}~R_{Schwarzchild}$ (of the central black hole) and about
1/15 the radius of the ionized disk.

\subsection{\it{Variability}}

Variability is a fundamental property of active galactic nuclei (e.g.,
\cite{kro99}) and has been observed in the optical continuum of the M87
nucleus (\cite{tsv98a}).  Variability at optical wavelengths typically
implies variability in the ultraviolet continuum as well.  
Photoionization can be an important mechanism for producing line emission.
If the photoionized gas is ionization bounded, then an increase in
the ionizing flux will lead to stronger Ly$\alpha$ emission.  This is
compatible with the M87 nucleus being quite transparent to Ly$\alpha$
photons (as we discussed in \S5) if the emitting gas is concentrated in
dense clouds that have a sufficiently low covering factor.
Furthermore, due to the travel time of photons from the central source
to the line emitting gas, the change in the continuum will be observed
first and change in line emissivity will be delayed (e.g.,
\cite{hor99}).  The increase in Ly$\alpha$ emission need not be related
to an increase in the continuum flux. It could also be a result of some
variability in the accretion flow that causes some excess gas to be
excited (either by photoionization or by shocks).  In any case, line
emission variability in the M87 nucleus is in itself not a surprising
phenomenon.

The shortest time scale for continuum variability observed by
\cite{tsv98a} was about a month.  In their set of 6 observations there
was only one pair separated by less than a week (the separation was 23
hours). Their sampling was not sufficient to exclude the possibility of
5 day variations.  Only continued monitoring of M87 can show reliably
whether variability on a 5 day time scale occurs at all, and if so how
common the occurrence is.

If we have serendipitously observed intrinsic variation in the
Ly$\alpha$ flux on a time scale $\tau \sim 5$ days, then the
observations set a limit on the size of the emitting region.  The
limiting size is $c\tau \sim 1.3\times10^{16}$ cm (0.005 pc).  This is
equivalent to about 100 $R_{Schwarzchild}$.


\section{Summary}

The spectrum of the M87 nucleus between 1140~\AA\ and 1606~\AA\ (with
250 km~s$^{-1}$ velocity resolution) shows continuum emission
consistent with previous observations and strong Ly$\alpha$ emission
with a large velocity width ($\sim$3000 km~s$^{-1}$).  The \ion{C}{4}
$\lambda$1549 line is marginally detected.  From the strength of the
Ly$\alpha$ and limit on the \ion{C}{4} emission we infer that the
nuclear line spectrum is different from the line spectrum of the
Keplerian disk presented by D97.  In particular the strong Ly$\alpha$
(and correspondingly, the weak \ion{C}{4}) could be due to the presence
of photoionized gas in addition to the shocked gas invoked by D97 to
explain the disk spectrum.  It could also be due to slower shocks that
do not produce \ion{C}{4} emission.

The strength of the observed Ly$\alpha$ also allows us to place some
rough limits on the amount and distribution of the absorbing gas
present within M87.  If the hydrogen column of the gas giving rise to
the observed metal lines (in absorption) is greater than about
10$^{18}$ cm$^{-2}$ then it cannot cover more than a small fraction of
the Ly$\alpha$ emitting region.

Our observations show that the Ly$\alpha$ line can be used effectively
to probe the kinematics of the central region of M87.  Spectra taken 5
days apart reveal a 60\% increase in the Ly$\alpha$ flux, but no change
in the continuum.  The Ly$\alpha$ flux difference is either due to {\it
(i)} an offset in the aperture position between the two observations or
{\it (ii)} intrinsic variability of the source.  To distinguish between
these two possibilities additional spectra with high pointing accuracy
($\le$ 0\farcs05) and sufficiently frequent time sampling (order of
days) will be needed.  As pointed out by \cite{tsv98a}, such
observations would have to be done with space-based telescopes since
the effects from background stars and the jet make it difficult to
isolate nuclear emission even in the best of seeing conditions from the
ground.

The kinematics of the central region of M87 could be mapped in detail
using the high spatial and spectral resolution capabilities of the {\it
Space Telescope Imaging Spectrograph} (STIS) on board the HST.  A time
series of long slit observations would allow one to distinguish between
the competing possibilities for the observed variation in the
Ly$\alpha$ emission profiles (\S6) as well as to study the metal-line
abundances and distribution of gas in the nuclear region.  Such
emission and absorption information could potentially be of great
importance for understanding the properties of material in the nuclear
disk and the physical processes related to the central black hole.


\acknowledgments
We thank Zlatan Tsvetanov for several discussions and Ray Lucas for
assistance with tracking down the epoch 1 pointing error.  We thank
the anonymous referee for useful suggestions.  We also
thank the STScI help desk for promptly answering questions about some details
of the observations.  We acknowledge using CADC, which is operated by
the National Research Council, Herzberg Institute of Astrophysics, Dominion
Astrophysical Observatory.
RS acknowledges support from grant GO-07289.01-96A from the Space
Telescope Science Institute.
KRS acknowledges support from NASA Long Term Space Astrophysics 
grant NAG5-3485.  CRC and KRS acknowledge support from NASA through 
STScI grant 05921-01-94A.



\newpage 

\figcaption{Continuum flux between $2.0\times10^{15}$~Hz and
$2.4\times10^{15}$~Hz (1500 \AA\~to 1250~\AA) from the nucleus of M87
(solid line) and from a position 1\arcsec\ away from the nucleus
(dotted line).  The continuum flux level is consistent with previous
measurements (see text).
            \label{CONT}}

\figcaption{The spectrum of the M87 nucleus (solid line) showing the
strong Ly$\alpha$ emission.  The line is broad and well separated from
the geocoronal Ly$\alpha$.  The dotted and dashed lines show two models
for the absorption due to Galactic hydrogen, plotted as transmittance
fractions.  They correspond to hydrogen columns 
of $3.4~\times~10^{20}$~cm$^{-2}$ and $2.1~\times~10^{20}$~cm$^{-2}$
respectively.
            \label{LYA}}

\figcaption{The spectrum of the M87 nucleus showing possible \ion{C}{4}
$\lambda$1549 emission.  The dotted line is the unsmoothed spectrum,
and the solid line is the spectrum rebinned over 4 pixels (about 1~\AA).
The dashed lines show estimated values for the \ion{C}{4} flux
based on previous observations of M87 (see \S4 for details).
            \label{C4}}

\figcaption{The dotted and dashed lines are two models for the absorption
due to hydrogen within M87, plotted as transmittance fractions.  As in 
Figure \protect\ref{LYA} these are overplotted on the observed M87 spectrum (solid
line). 
	    \label{ABS}}

\figcaption{Top Panel: Spectra of the M87 nucleus taken 5 days apart (epoch 2).
The integrated Ly$\alpha$ flux in the later observation is about 60\% higher than
in the earlier observation.  The change could be due to a slight offset
in the aperture locations between observations or to intrinsic
variability of the source.  The spectrum shown in Figures \protect\ref{LYA} 
and \protect\ref{ABS} is an average of these two spectra.
Bottom Panel: The ratio of epoch 2b flux to epoch 2a flux for the wavelength
region 1218 \AA\ to 1230 \AA.  Also shown are the observed fluxes on an
arbitrary scale.  The fluxes and the ratio plot have been binned over 4 pixels.
The horizontal bars represent the errors introduced due to the binning.
            \label{LYAVAR}}

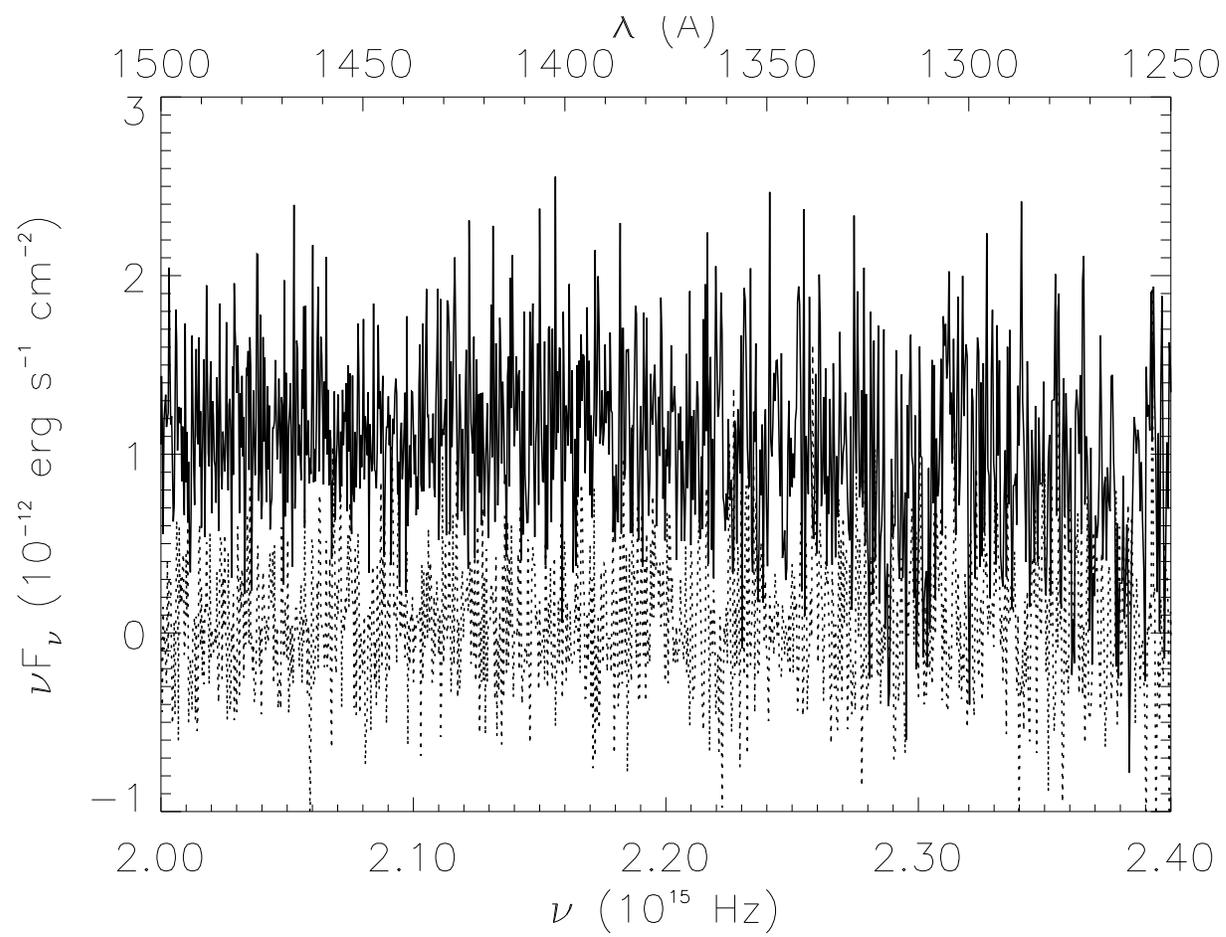
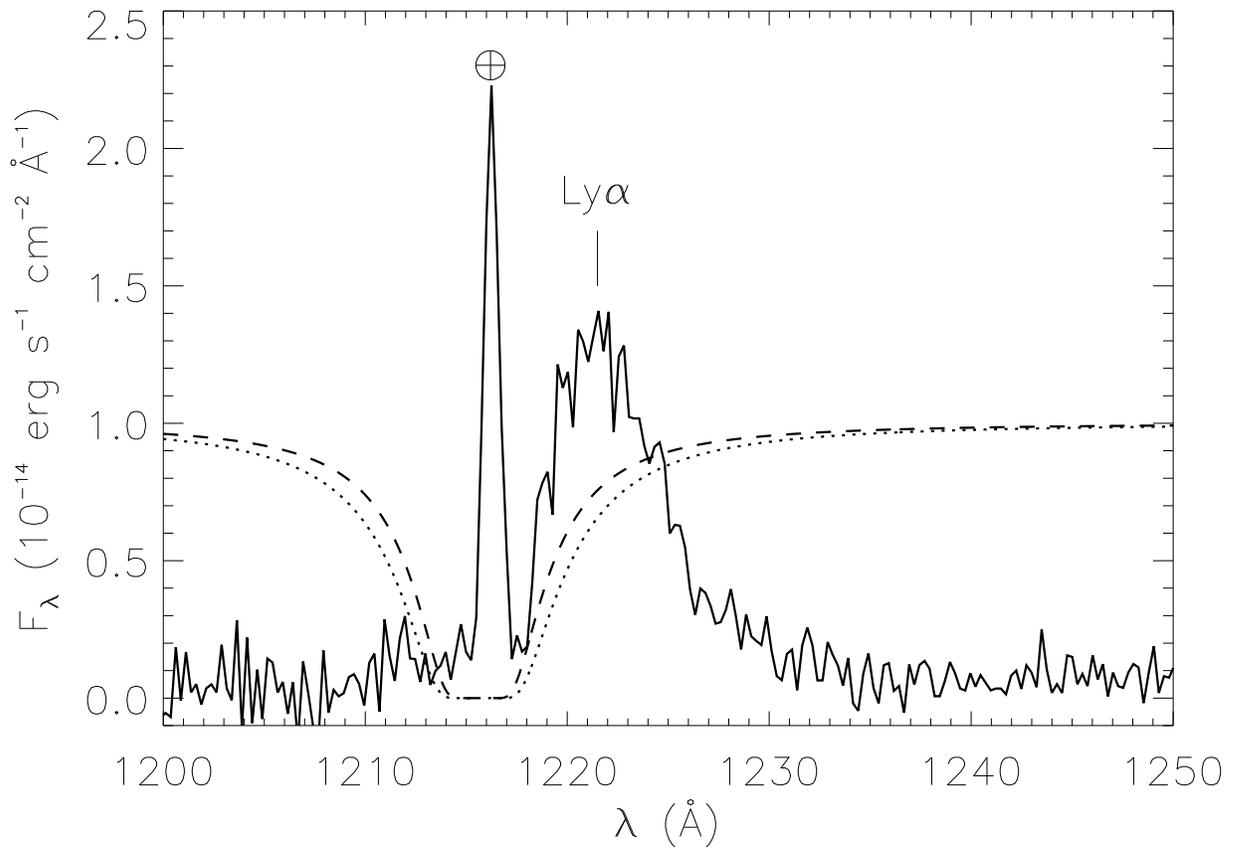
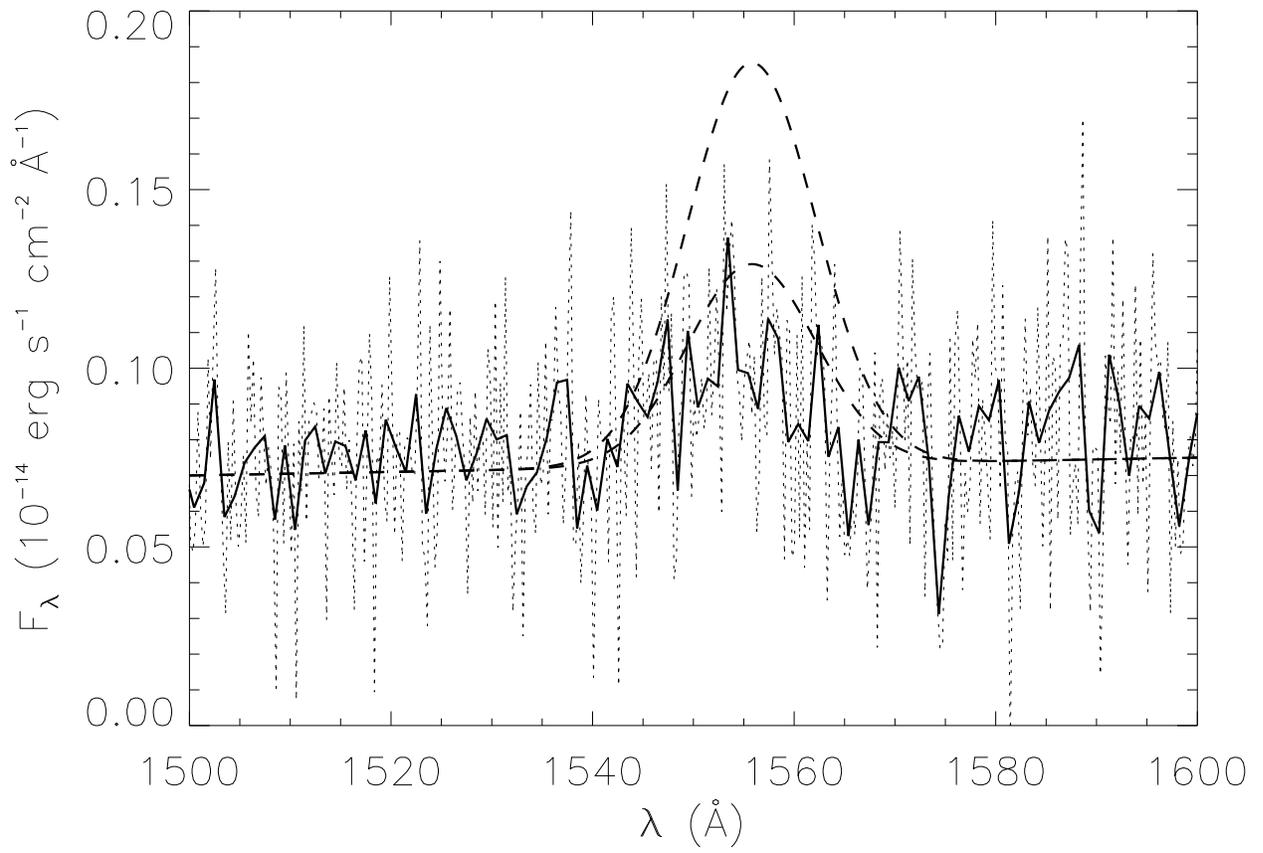
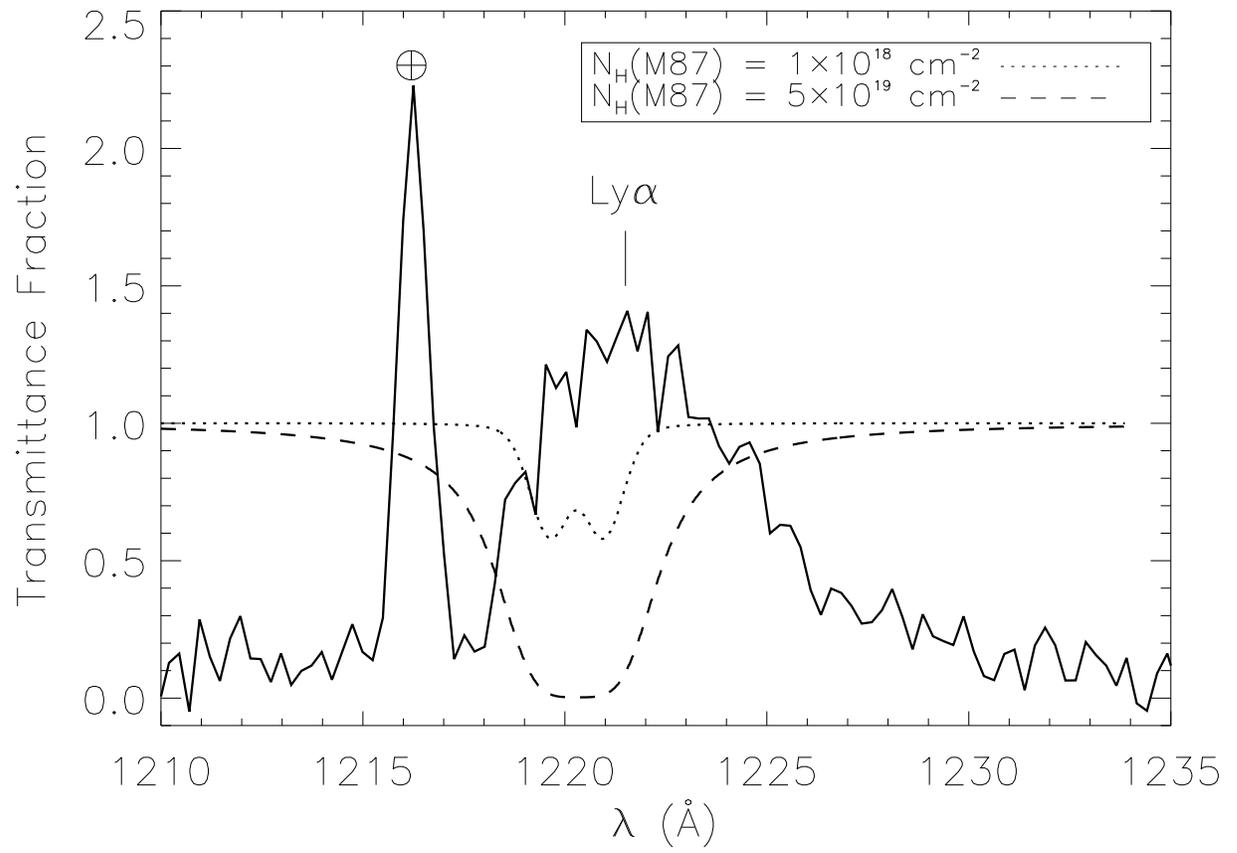
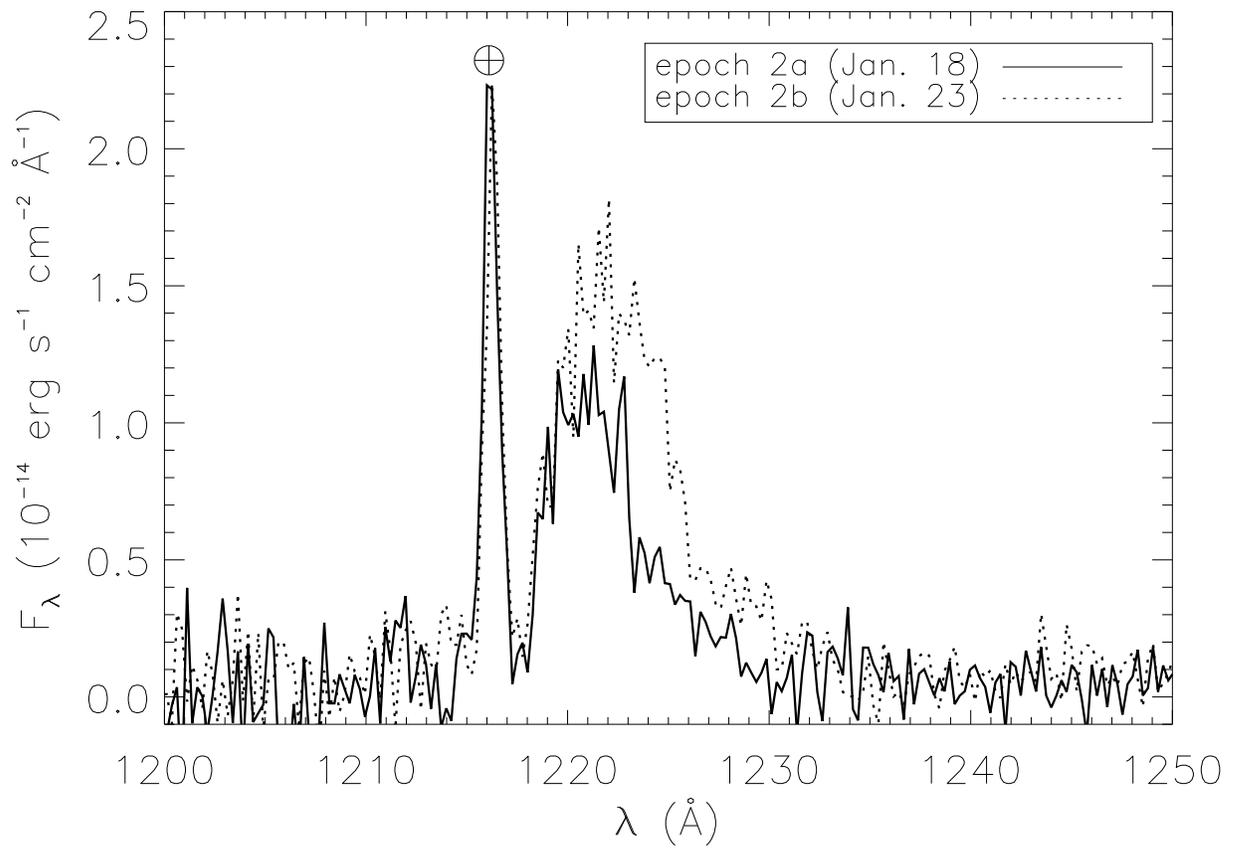
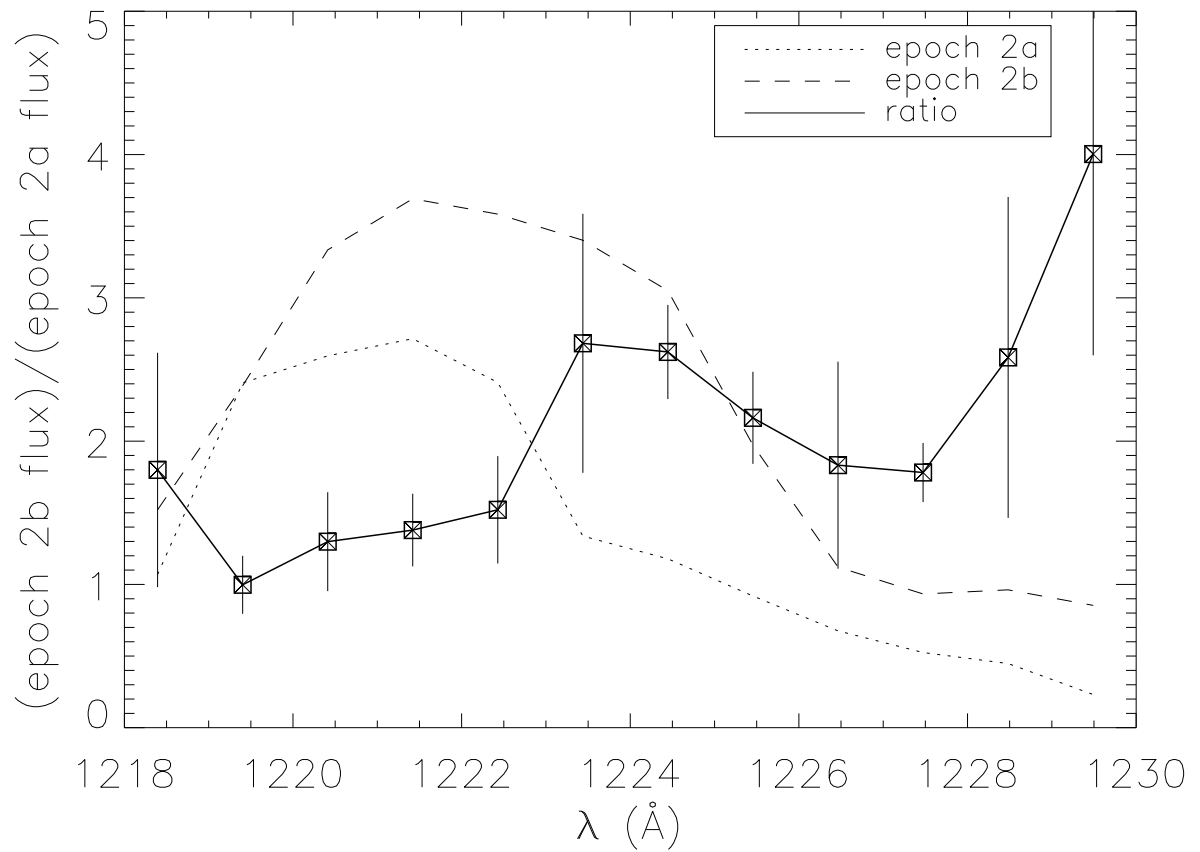
\end{document}